\documentclass{optica-article}     % document class for the Optica Journals
\journal{opticajournal}            % for journals or Optica Open
\articletype{Research Article}     % For research articles 
\usepackage{comment}
\usepackage{soul}
\usepackage{xfrac}

\begin{document}
% \linenumbers     % Turn off line numbering for Optica Open preprint submissions.

% Title ==================================================================== %
\title{Acoustoelectrically enhanced acousto-optic modulation in an integrated silicon nitride and thin film lithium niobate platform}
% \title{Other working title ideas}

% Authors ================================================================== %
\author{Matthew J. Storey\authormark{1,$\dag$,*}, John H. Dallyn\authormark{1,2,$\dag$},Kiyan Hocek\authormark{3,$\dag$}, Michael Miller\authormark{1}, Peter T. Rakich\authormark{4},Scott A. Diddams\authormark{3},
 Nils T. Otterstrom\authormark{1} and Matt Eichenfield\authormark{1,3,*}}

\address{\authormark{1}Microsystems Engineering, Science and Applications, Sandia National Laboratories, P.O. Box 5800 Albuquerque, NM 87185, USA\\
\authormark{2}Department of Applied Physics and Material Sciences, Northern Arizona University, 1900 S Knoles Dr., Flagstaff, AZ 86011, USA\\
\authormark{3}Electrical, Computer \& Energy Engineering, University of Colorado, Boulder, CO 80309\\
\authormark{4} Department of Applied Physics, Yale University, New Haven, CT 06520, USA\\
\authormark{\dag The authors contributed equally to this work.}}

\email{\authormark{*}mjstore@sandia.gov, matt.eichenfield@colorado.edu} %% email address is required; see note below about the corresponding author designation

% ========================================================================== %
% -------------------------------------------------------------------------- %
%                                 Abstract                                   %
% -------------------------------------------------------------------------- %
% ========================================================================== %

% use {asbstract*} to suppress the copyright line. Copyright information will be added in production
\begin{abstract*}
Acoustoelectric interactions in piezoelectric-semiconductor heterostructures allow the propagation characteristics of microwave frequency phonons in piezoelectric media to be controlled and radically enhanced, providing electrically controllable phonon gain, large velocity tuning, isolation, and circulation, as well as extremely large electron-mediated phononic nonlinearities. Here, for the first time, we create such a piezoelectric-semiconductor heterostructure with lithium-niobate-on-insulator and InGaAs that also supports guided optical modes through the addition of a silicon nitride waveguide and modification of the acoustic materials to provide an optical lower cladding. We use this new architecture to demonstrate acoustoelectrically enhanced acousto-optic modulation, where 1 GHz phonons are piezoelectrically generated and acoustoelectrically amplified on-chip by up to 60 dB before impinging on the optical waveguide, providing pure phase modulation with a $V_\pi L$ figure-of-merit of 0.077 V-cm while only consuming 3.77 mW of DC electrical power to provide the amplification. We then consider future applications enabled by these functionalities and describe a novel tunable optical delay and an optoelectronic oscillator (OEO) analog---an   acoustoelectrically enhanced opto-acoustic oscillator (AE-OAO). We show that using Brillouin optomechanical transduction and acoustoelectrically lossless acoustic time delay, the AE-OAO could replace kilometers of optical fiber delay used in OEOs but on a single, centimeter-scale chip.
\end{abstract*}

% ========================================================================== %
% -------------------------------------------------------------------------- %
%                               Introduction                                 %
% -------------------------------------------------------------------------- %
% ========================================================================== %

\section{Introduction}

Over the past two decades, a branch of microwave photonic (MWP) systems utilizing RF acoustics and optomechanics \cite{capmany2007microwave,marpaung2019integrated} has developed that offers a powerful platform for engineering RF–optical links. By coupling light to microwave-frequency acoustic waves in cavity and traveling-wave topologies, the field of acoustic-based MWP has demonstrated tunable delay \cite{aryanfar2017chip,mckay2020integrated}, narrowband filters \cite{gertler2022narrowband}, broadband modulators \cite{yao2002brillouin,erdil2024wideband,tadesse2014sub,zhang2024integrated,shao2020integrated}, large optical carrier frequency shifts \cite{wang2013optical}, and optical nonreciprocity that allows for isolation and circulation \cite{tian2021magnetic,sohn2021electrically,peterson2018synthetic}. Yet these demonstrations remain fundamentally constrained: the degree of nonreciprocity is limited, amplification of RF sidebands requires them to be transduced onto high-power lasers, the demonstrated means of RF to optical transduction are difficult to implement, and in all cases the functionalities lack the kind of straightforward voltage-controlled programmability that defines practical RF front ends. As a result, despite remarkable advances, optomechanics alone have not been able to provide the full suite of active, reconfigurable operations demanded in MWP RF–optical links.

In parallel, acoustoelectric (AE) systems \cite{white1962amplification,carleton1965ultrasonic,white1967surface} have emerged as a complementary platform with precisely these missing capabilities. Built from piezoelectric semiconductor heterostructures, they enable direct voltage control of processes that allow the manipulation of all phononic properties and have demonstrated reconfigurable high-gain, low-noise, and low-power amplification \cite{hackett2019amp, hackett2021towards, hackett2023non}; orders of magnitude more nonreciprocal transmission for isolation and circulation than their optomechanical (OM) counterparts \cite{hackett2023non}; direct RF switching\cite{storey2021acoustoelectric}; simultaneously low-power-consumption and high-output-power phonon source generation\cite{wendt2025electrically}; phase shifters \cite{crowley1977acoustoelectric,pedros2010voltage,yamagata2022surface}; and efficient three- and four-wave mixing processes that allow modification of the RF spectral content \cite{hackett2024giant}.

    The hybridization of optomechanical and acoustoelectric systems thus offers a route past the long-standing limitations of MWP. Optomechanics provides broadband, low-power microwave–optical transduction and naturally tunable band-selectivity through optical resonances, and acoustoelectrics can then provide the active, nonlinear, and voltage-controlled RF functionalities directly on those transduced phonons that are otherwise missing or only weakly expressed in optical systems. By combining the two within a common piezoelectric semiconductor platform that also allows for low-loss optical waveguiding, it becomes possible to envision a monolithic architecture in which RF-acoustic signals are processed with amplification, isolation, switching, and multi-wave mixing processes in the RF domain, and then up-converted to light \cite{zhou2024electrically}. Such an approach promises to overcome the practical barriers of previous RF–photonics efforts, enabling low-power, agile, and multifunctional RF–optical links for classical communications and signal processing. 

\begin{figure*}[ht]
    \centering
    \includegraphics[width=1\linewidth]{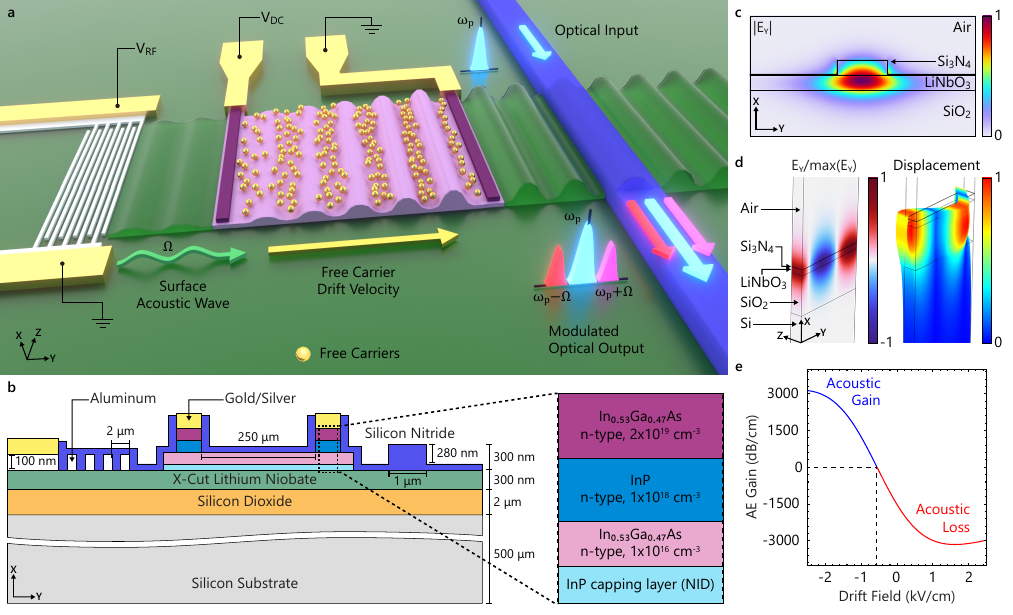}
    \caption{(a) Device operation schematic. A RF voltage $V_{RF}$ is applied to an IDT which launches an acoustic wave. This surface acoustic wave with frequency $\Omega$ encounters a InGaAs semiconductor layer. A DC voltage $V_{DC}$ is applied in such a way to drive the free carriers in the semiconductor in the direction of acoustic wave propagation. Acoustoelectric coupling between the acoustic wave and free carriers results in acoustic gain. The acoustic wave exits the acoustoelectic interaction region and encounters a silicon nitride optical waveguide with optical signal at frequency $\omega_0$. After modulation, the light exits with frequencies $\omega_0$ and $\omega_0\pm\Omega$. (b) Cross section of the acoustoelectrically enhanced acousto-optic modulator. (c) Simulation of the optical mode in the silicon nitrade waveguide. (d) Electric field (right) and acoustic wave (left) simulations in the Si, SiO$_2$, LiNbO$_2$, Si$_3$N$_4$ stack. (e) Example of an acoustoelectric gain curve. Depending on electric field orientation there is acoustic gain or acoustic loss. Added loss inherent to acoustocelectric material (acoustoelectric drag) shifts the 0 dB/cm AE gain point away from 0 drift field, indicated by the dashed lines.}
    \label{fig:op-principle}
\end{figure*}

Here, we take a first step toward this vision by designing a modular acoustoelectrically enhanced optomechanical (MAEO) platform. This is accomplished through an optical-waveguide-integrated, piezoelectric-semiconductor heterostructure. As a first demonstration, we show reconfigurable acoustoelectric amplification of RF signals transduced onto the chip as phonons and then transduced onto an optical carrier by optomechanical modulation. Our device consists of a lithium niobate electromechanical transducer, an InGaAs-LiN$\rm O_3$ acoustoelectric gain region, and a silicon nitride optical waveguide for optomechanical readout. We demonstrate record-high (>60 dB) reconfigurable acoustic gain and use this process to greatly enhance optomechanical phase modulation imprinted on a telecom-band carrier. This enhancement corresponds to a 1000$\times$ reduction in the effective $V_{\rm \pi}$ of the optical modulator, resulting in an acoustoelectrically improved $V_{\rm \pi}L=0.077$ $\rm V$ $\rm cm$. This platform represents the first monolithic architecture that enables fully independent control over the optical carrier and RF spectral content, permitting complementary signal processing capabilities of optical, RF, and acoustic domains.

% ========================================================================== %
% -------------------------------------------------------------------------- %
%                                 Results                                    %
% -------------------------------------------------------------------------- %
% ========================================================================== %

\section{Results}

% ========================================================================== %
%                               Device Design                                %
% ========================================================================== %

\subsection{Device design}

Full demonstrations of acoustoelectric enhancement of a stimulated optomechanical Brillouin scattering process place exceedingly stringent demands on the device, platform, and material properties. For instance, the platform must support co-linear and co-localized low-loss optical and acoustic modes, have a significant piezoelectric response, and have the proper charge carrier properties within the appropriate semiconductor region. Given this device complexity, we set out to create the MAEO platform that permits independent design freedom for the acoustoelectric and optical components of the device. While this precludes acoustoelectric enhancement of a stimulated (or self-reinforcing) optomechanical process (such as SBS), this MAEO platform is sufficient for demonstrations of acoustoelectrically enhanced acousto-optic modulation processes.

A schematic of the platform and device concept is presented in Fig. \ref{fig:op-principle}a. A surface acoustic wave (SAW) delay path, whose initial conditions are defined by an interdigitated transducer (IDT), is intersected perpendicularly by an optical waveguide. Acousto-optic phase modulation is imprinted on an optical carrier through a form of intra-modal Brillouin scattering. Thanks to the vanishingly small acoustic wavevector along the direction of optical propagation (i.e., the projection along the x axis is zero), the SAW permits simultaneously Stokes and anti-Stokes scattering, resulting in nearly ideal optical phase modulation. Along the SAW path, an acoustoelectric amplifier region is formed by a thin semiconductor layer that sits atop the piezoelectric device layer. An applied voltage and resulting electric field in this region creates a drift current corresponding to a charge carrier velocity that can exceed that of the SAW, producing acoustoelectric phonon gain that consequently enhances the acousto-optic scattering process.

% ========================================================================== %
%                                Fabrication                                 %
% ========================================================================== %

\subsection{Device fabrication}

\begin{figure*}[ht]
\centering
\includegraphics[width=1\linewidth]{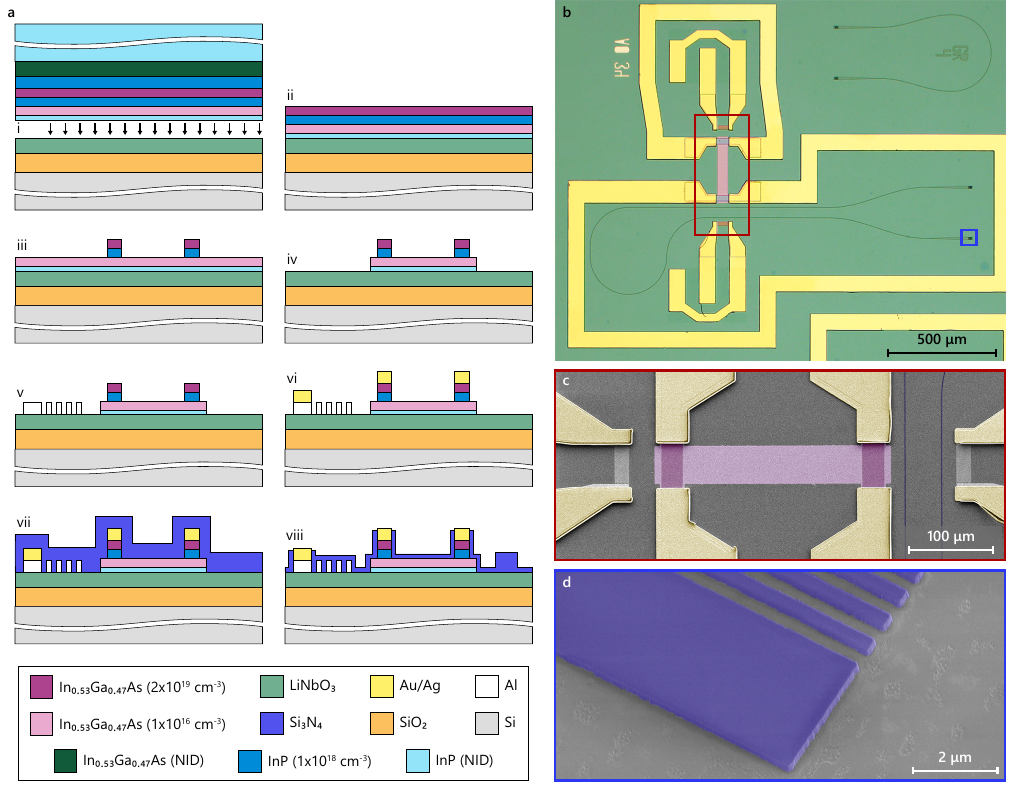}
\caption{(a) Fabrication flow for MAEO platform. (b) optical microscope image of the fabricated acousto-optic amplifier. (c) SEM of silicon nitride optical grating coupler while (d) is an image of the interdigitated transducer (IDT) (left), acoustoelectric gain region (middle; purple), a loop-back optical waveguide structure for optomechanical readout (right), and an output IDT (far right).}
\label{fig:fab-process}
\end{figure*}

The MAEO device consists of a lithium niobate IDT, InGaAs-LiN$\rm O_3$ acoustoelectric heterostructre, and silicon nitride optical waveguide, as shown top-down in Fig. \ref{fig:fab-process}a-c. The fabrication flow is illustrated in Fig. \ref{fig:fab-process}a. (i) Fabrication begins by bonding the metal organic chemical vapor deposition grown InGaAs/InP heterostructure to the LN-O-Si wafer. (ii) Successive wet etching is employed to thin the InP and InGaAs layers using  a HCl/H3PO4 mixture and dilute piranha solution respectively. (iii) After etching of the InP seeding wafer, the NID InGaAs buffer layer, and the InP etch stop layer, electron beam lithography (EBL) using a negative resist in conjunction with a successive wet etch described above is used to define the contacts region of the epitaxial film. (iv) This lithography to etch process is repeated once more to define the amplification layer of the film, with the wet etch landing on the LiNbO3 thin film. (v) EBL was once again employed to pattern the electron beam evaporated aluminum inter digitated transducer (IDT) structure using a standard PMMA Bi-Layer liftoff recipe. (vi) A thick PMMA Bi-Layer recipe once again in conjunction with electron beam evaporation was used in order to facilitate liftoff of the thick Au/Ag metal layer, leaving the Au/Ag at the top of the contact stacks.
(vii) A 300nm thick conformal layer of Si3N4 was deposited using plasma enhanced chemical vapor deposition (PECVD) to not only passivate the III-V structures, but also act as the waveguiding layer. (vii) EBL was once more used to pattern the optical waveguiding structures this time with a negative resist. Reactive Ion Etching (RIE) using a standard C-F and argon chemistry. The etch is intentionally stopped short of landing on the LN thin film in order to keep the III-V layer passivated. A last round of EBL-RIE etching is used in order to fully open access to the Au/Ag layer.

% ========================================================================== %
%    Measurement of acoustoelectrically enhanced acousto-optic modulation    %
% ========================================================================== %
\subsection{Measurement of acoustoelectrically enhanced acousto-optic modulation}

\begin{figure*}[ht]
\centering
\includegraphics[width=1\linewidth]{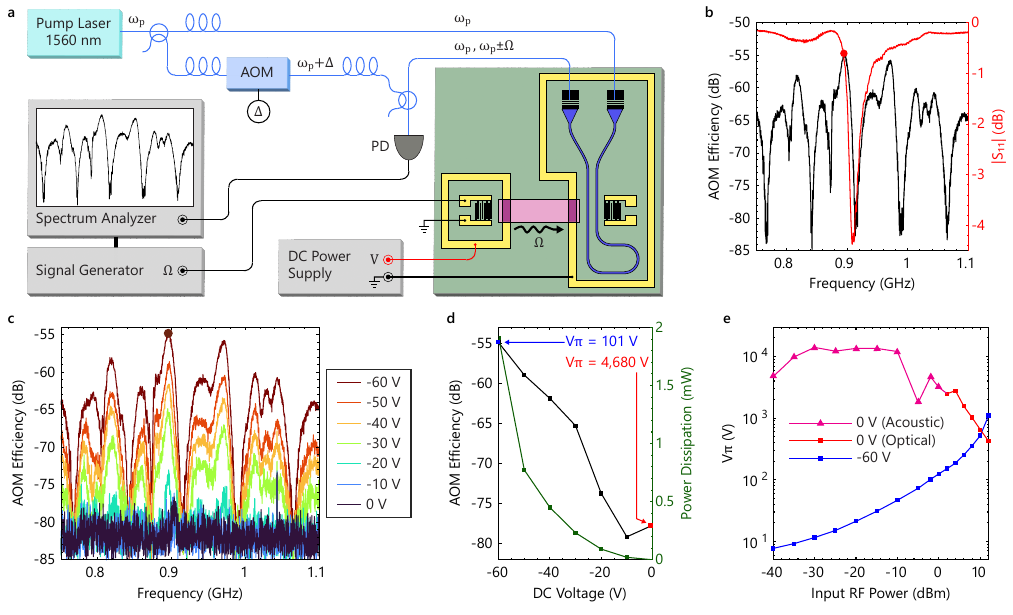}
\caption{Acoustoelectrically enhanced acousto-optic modulator results. (a) Setup of the heterodyne measurement of the modulated optical signal. A pump laser at frequency $\omega_p$ is split into two paths. One path is modulate by an acousto-optic modulator (AOM) driven by a local oscillator at frequency $\Delta$. The other path couples into the on-chip modulator device. A signal generator sweeps through IDT-generated acoustic wave frequency $\Omega$. A DC power supply drives the current to produce acoustoelectric gain in the purple interaction region. The modulated optical signal $\omega_p\pm\Omega$ is coupled out of the chip where it is mixed the local oscillator arm into a photodetector. The photodetector is readout to a spectrum analyzer. (b) Acoustoelectrically enhanced acousto-optic modulator optical efficiency in black. The applied DC voltage in -60 V. In red, the S$_{11}$ IDT transmission frequency sweep measured using a IDT readout of the acoustic wave into a network analyzer. Demonstrated a filtering effect due to the acoustic wave passing through the waveguide twice. (c) Acoustoelectrically enhanced acousto-optic modulator optical efficiency at applied voltages from 0 V to -60 V. (d) Plot of the acoustoelectrically enhanced acousto-optic modulator peak optical efficiency at various voltages. The $V_\pi$ calculations for 0 V and -60 V were 4680 V and 101 V, a 46x improvement. (e) Plot of the $V_\pi$ for -60 V (blue squares) and 0 V (red squares and magenta triangles) for various input RF powers. Red squares indicate $V_\pi$ values calculated from modulated optical signals measured with the setup in part (a). At RF input powers $\leq 0$ dBm (magenta triangles), the optical signal is below the spectrum analyzer's noise floor. Direct measurements of the acoustic wave using a VNA through the second IDT is used to interpolate the $V_\pi$ values.}
\label{fig:AO-results}
\end{figure*}

To measure the acoustoelectrically enhanced acousto-optic modulation, we use the heterodyne measurement setup illustrated schematically in Fig. \ref{fig:AO-results}a. Starting with a probe laser operating at $\omega_p=1560$ nm, we divide the beam into two paths, the signal arm and the reference arm. In the signal arm, the light is surface coupled into an on-chip silicon nitride waveguide. While in the waveguide, the light encounters amplified surface acoustic waves with frequency $\Omega$ in the region between the epitaxial semiconductor layer (purple) and the output IDT. These acoustic waves cause periodic photoelastic variations in the waveguide material index, scattering light into sidebands with frequencies $\omega_p\pm n\Omega$, where $n$ is the sideband order. The modulated light is subsequently coupled off chip and combined with the blue-shifted optical local oscillator (LO) light from the reference arm for heterodyne spectroscopy. The optical LO is synthesized using a commercial acousto-optic frequency shifter, which blueshifts the light by $\Delta=43$ MHz, resulting in a reference frequency $\omega_{\rm LO}=\omega_p+\Delta$. The combination of the modulated signal sidebands and optical LO result in optical beat notes at microwave frequencies $|n|\Omega\pm \Delta$ for the red and blue sidebands, respectively, which are measured with an RF spectrum analyzer (SA), as depicted in Fig. \ref{fig:AO-results}a.

The acoustoelectrically enhanced acousto-optic scattering process is measured using a form of driven frequency-resolved heterodyne spectroscopy. The signal generator, which supplies the RF signal to the input IDT, is synchronized and controlled by the RF SA. An example of the measured spectrum is given in Fig. \ref{fig:AO-results}b. Additionally, the S$_{11}$ IDT transmission frequency sweep measured using a IDT readout of the acoustic wave into a network analyzer is in red. We observe that the max transduction of the SHSAW occurs at the same frequency as severe dip in the optical signal. This occurs due to an unintentional filtering effect from the acoustic wave passing through the waveguide twice, in which the modulation from the acoustic wave is canceled out\cite{katzman2021surface,munk2019surface}. By chance, this resulted in a suppression of the peak optical modulation frequency, obscuring the full performance of these devices. 
 
The RF drive frequency $\Omega$ is swept through the IDT resonant response, and the resulting acousto-optic modulation response is measured for a range applied DC biases to the acoustoelectric amplifier region. In Fig. \ref{fig:AO-results}c, we plot the acousto-optic scattering efficiency, the acousto-optic modulation sideband power divided by the carrier power ($P_1/P_0$). The input RF power was -2 dBm. We see an increase in efficiency as higher DC voltages are applied, increasing the free carrier drift velocity and consequently enhancing the acoustoelectric coupling. We apply a negative voltage in order to bias the free carriers in the direction of acoustic wave propagation. In Fig. \ref{fig:AO-results}d, we plot the peak optical signal at frequency 896 MHz for various applied DC voltages. At a low voltage (-10 V), there is no improvement in the AOM efficiency, likely due to acoustoelectric drag (Fig. \ref{fig:op-principle}e). At -20 V and above, acoustoelectric gain overcomes the added loss and we observe an AOM efficiency enhancement of 24 dB with a -60 V DC bias. 

We characterize the acoustoelectric enhancement of the optomechanical process by considering the acoustoelectrically dependent half wave voltage $V_\pi$ of the acousto-optic phase modulator under the small modulation approximation  \cite{gao2024compact,ren2019integrated} (details in Supplement)
\begin{align}\label{Vpi}
    V_\pi\approx\frac{\pi V_{RF}}{\sqrt{\eta}}.
\end{align}
In Fig. \ref{fig:AO-results}d, we calculate the $V_\pi$ for 0 V and -60 V and see a 46$\times$ improvement in our modulator performance due to the acoustoelectric gain. In Fig. \ref{fig:AO-results}e, we plot $V_\pi$ as a function of input RF power from -40 dBm to 12 dBm for DC voltages of 0 (red squares and magenta triangles) and -60 V (blue squares). With a 0 V DC bias, RF powers less than 0 dBm (magenta triangles) produced an acousto-optic response below the noise floor of the spectrum analyzer. Therefore, we used an vector network analyzer (VNA) to directly measure the acoustic wave using the IDT on the right side (according to Fig. \ref{fig:AO-results}a) of the optical waveguides. We interpolate the $V_\pi$ values by assuming that the ratio of the VNA measurements for 0 V and - 60 V is equivalent to the ratio of the optical measurements, giving an approximation for the 0 V efficiencies. Due to the destructive interference of the optical signals near the most efficient frequency of operation for the IDT, our AOM efficiency could be improved by 14 dB by removing this interference. At larger input powers, the AE $V_\pi$ improvement significantly decays due to saturation of the acoustoelectric effect \cite{hackett2021towards}. However, our acoustoelectrically enhanced device performs very well at low RF power inputs, reaching a $V_\pi L$ of 0.077 V-cm at a -40 dBm input with only 3.77 mW of dissipated power and a modulator length of 100 $\mu$m (50 $\mu$m IDT aperture times two optical waveguide passes). This is a 1000$\times$ reduction of the effective halfwave voltage compared to 0 V at the same input power. 

% ========================================================================== %
% -------------------------------------------------------------------------- %
%                        Outlook and Discussion                              %
% -------------------------------------------------------------------------- %
% ========================================================================== %

\section{Outlook and Discussion}
\begin{figure*}[ht]
    \centering
    \includegraphics[width=1\linewidth]{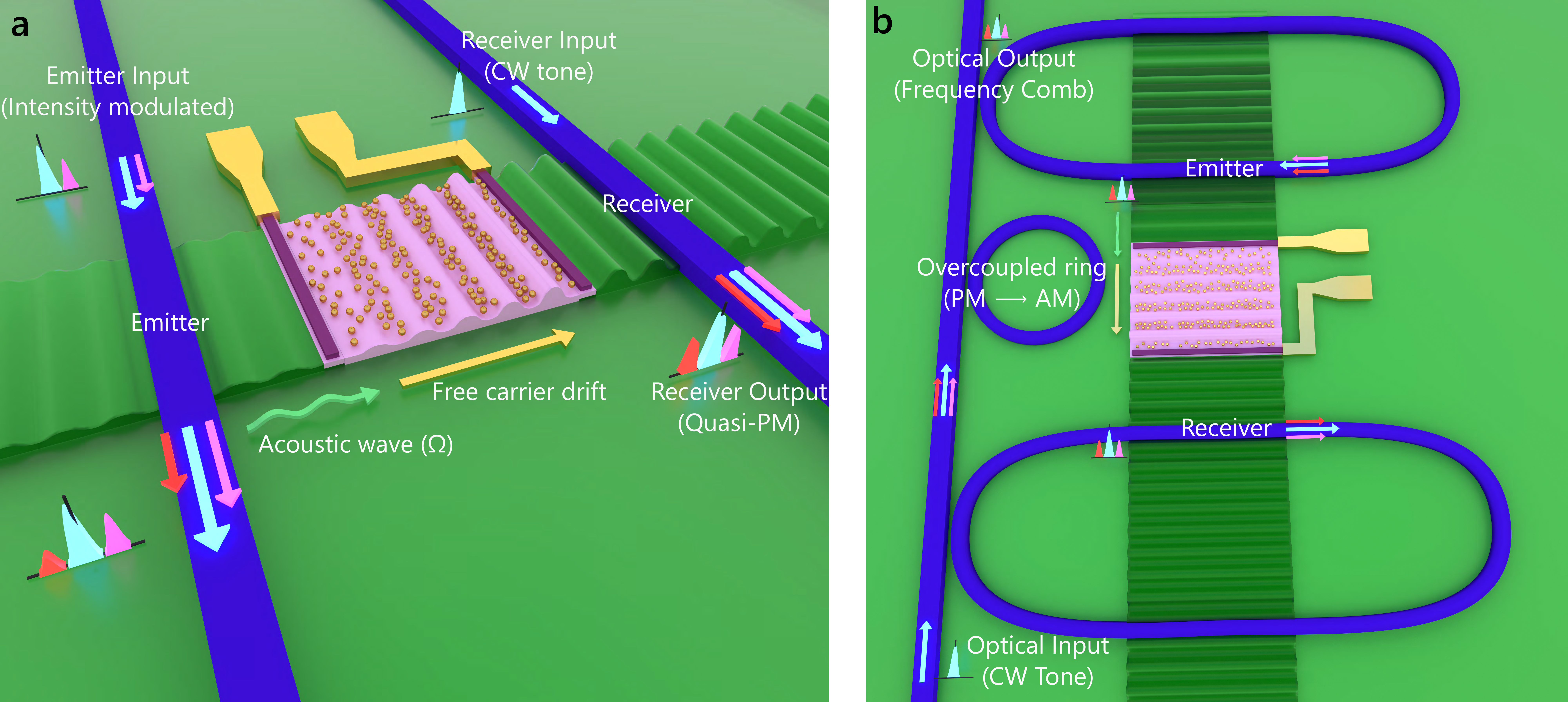}
    \caption{(a) PPER operation. In contrast to the AOM discussed above, an RF acoustic wave is generated by an intensity modulated signal in the emitter waveguide. This allows for reconfigurable optical signal processing with acoustic amplifiers, such as tunable optical delay or a hybrid acoustic/optic oscillator (AE-OAO) shown in (b). The AE-OAO operates analogous to an OEO, with the electronic path, including photodetection, optical modulation, amplification, and optical delay functionally replaced by an acoustic delay line that interacts through forward Brillouin scattering with the optical signal.}
    \label{fig:AE-OAO-PPER}
\end{figure*}

Looking forward, we note that many microwave photonic systems require optoelectronic transduction, photodetectors,  RF amplifiers, and optical delay lines. %include citations,
The platform we have demonstrated here is a first step towards combining Brillouin optomechanical processes and acousto-electric components that can replace and potentially even out-perform many of these components and systems. Building on what we have demonstrated here using Brillouin optomechanical transduction between a waveguide and an approximately perpendicularly propagating acoustic wave, we first note that this transduction can be made to go in both directions and extremely efficient using so-called phonon-photon emitter receivers (PPER) \cite{Shin_Cox_Jarecki_Starbuck_Wang_Rakich_2015}. The PPER structure consists of two optical waveguides of length $L$ and separated by a distance $d$ between them. An amplitude modulated optical signal generates a phonon in the emitter waveguide through Brillouin optomechanical interactions and the receiver can transduce signals back to the optical domain with a simple probe beam being phase-modulated by the phonon field from the emitter. This immediately opens the possibility of placing acoustoelectric signal processing components between the emitter and receiver, creating an AE-PPER paradigm that is quite powerful. 

The simplest case of the AE-PPER would be a continuously tunable optical delay line amplifier. This delay line amplifier can reconfigurably modify both the gain and velocity of the phonons. As demonstrated experimentally, here, the gains can be extremely large, simply compensate for propagation loss, or even be made to attenuate and quench the phonon. The velocity shifts are also tunable and enormous, capable of tuning over a range of order $(1/2)k^2$ and can be made \textit{lossless} with the gain. As an example, in thin-film LiNbO$_3$ with velocities of order $~$4500 m/s and $k^2$ that can reach $~$40\% \cite{896145}, a 20 micron AE-PPER phononic delay line as shown in Fig. \ref{fig:AE-OAO-PPER}a) could provide 4.4 ns of fixed delay and 1 nanosecond of \textit{continuously voltage-tunable delay}.

Going a step further, we now describe an AE-PPER \textit{oscillator}, which we call an acoustoelecrically enhanced opto-acoustic oscillator (AE-OAO). The AE-OAO is analogous to the opto-electronic oscillator (OEO), which generates a frequency comb (mode-locked laser) and acts as an RF oscillator, but the AE-OAO can be made on a \textit{single chip} and does not require an RF amplifier, RF phase-shifter, photodetector, optical amplifier, or electro-optic modulator \cite{Xie_Sun_Peng_Zhang_Guo_Zhu_Hu_Chen_2014}. Instead, the entire electronic/RF signal chain of the OEO is replaced by a phononic path and amplifier on the chip, with transduction between photons and phonons, delay, and gain, provided by an AE-PPER. Perhaps even more importantly, ultra-low phase-noise OEO require long, often kilometer-scale fiber delay to reduce the phase noise by increasing the round-trip time of the feedback signal \cite{Yao_Maleki_1996}. However, the acoustic delay line greatly reduces the physical length due to the $\sim 10^4-10^5$ reduction in velocity compared to the speed of light, allowing comparable round-trip time delays on a single chip, and the amplifying functionality can provide anywhere from lossless propagation to large gain.

As shown in Figure \ref{fig:AE-OAO-PPER}b), the receiver waveguide forms an optical resonator with a free-spectral range at the desired oscillator frequency increasing the effective modulation depth. The light in the racetrack couples back out into the bus waveguide and passes near an overcoupled ring resonator, converting the phase-modulated signal into intensity modulation at the acoustic frequency. The intensity modulated signal circulates in a separate resonator identical to the receiver side, where the intensity modulation excites, via optical forces, an acoustic wave that travels through the gain region and reseeds the phase modulation. When the round trip gain of the signal exceeds unity, $|G_s| > 1$, the device will oscillate at a frequency determined by the allowed optical frequencies of the racetrack resonator. Further, the round-trip resonance condition requires $\arg{G_{s}} = 2\pi n$. The small-signal gain condition can be shown to be (see Supplemental Information)
\begin{align}
    G_{s} =  \frac{i|g|^2 \eta_{p\rightarrow m} \chi_m G_{ae} L_{pm} }{ v^2 \hbar \omega }  \frac{r^2\rho\left(t(1-\rho) - r^2 \right)}{(1-\rho)^3} P_{laser}
\end{align}

We have presented an acoustoelectrically enhanced silicon nitride acousto-optic modulator on a lithium niobate-based platform with heterogeneously integrated InGaAs. This work represents the first demonstration of acoustoelectrically enhanced acousto-optics, and more generally, the first results showing a clear AE enhancement of an optomechanical (OM) interaction. While the AE and OM dynamics are linked only unidirectionally in this experiment (i.e., no reverse transduction), this AE-enhanced modulator has clear technological impact and is an important first step toward AE-enhanced stimulated optomechanical processes ranging from OM oscillation and lasing to Brillouin-based filtering and amplification. Apart from its novelty in connecting two distinct phenomena for the first time, this non-optimized AE-enhanced modulator is actually itself an interesting high-performance component. The AE properties alone are exquisite, enabling record-large RF acoustic non-reciprocity (>60 dB) in continuous wave operation. For an RF input of -40 dBm and a applied voltage of -60 V, our devices achieve a resonant modulator figure of merit $V_\pi L$ of 0.038 V-cm, comparing very favorably with state-of-the-art lithium niobate platforms\cite{qi2020integrated}. Furthermore, the performance could be improved with straightforward  modifications, such as correcting the optical waveguide separation to prevent destructive acousto-optic interference on resonance \cite{munk2019surface}, the addition of many constructively-interfering acousto-optic waveguide segments to further improve responsivity \cite{kittlaus2021electrically,kenning2025broadband}, and the development of isolation vias that shield regions where AO modulation is not desired \cite{Hassanien2021LNrel}. The last improvement is especially intriguing as it could enable a new capability for acousto-optics not accessible with traditional methods: the ability to rapidly control phonon attenuation would permit acousto-optic switches and shutters that are not limited in speed by the passive acoustic properties of the system, such as ring down or acoustic transit time.

% ========================================================================== %
% -------------------------------------------------------------------------- %
%                                 Methods                                    %
% -------------------------------------------------------------------------- %
% ========================================================================== %

\section{Methods}
The acousto-optic modulator measurements are taken using heterodyne measurement. Pump laser light is split into two paths. One path is coupled on and off chip via grating couplers. The other path is modulated at frequency $\Delta$ using an AOM, then mixed the the light coupled out of the chip onto a photodetector. Meanwhile, an IDT on chip is driven using an RF signal generator and a drift current across the InGaAs is provided by a DC power supply.

% ========================================================================== %
% -------------------------------------------------------------------------- %
%                               Back Matter                                  %
% -------------------------------------------------------------------------- %
% ========================================================================== %

\begin{backmatter}

% ========================================================================== %
%                                 Funding                                    %
% ========================================================================== %

% \bmsection{Funding}
% Content in the funding section will be generated entirely from details submitted to Prism. Authors may add placeholder text in this section to assess length, but any text added to this section will be replaced during production and will display official funder names along with any grant numbers provided. If additional details about a funder are required, they may be added to the Acknowledgment, even if this duplicates some information in the funding section. For preprint submissions, please include funder names and grant numbers in the manuscript.

% ========================================================================== %
%                               Acknowledgment                               %
% ========================================================================== %

\bmsection{Acknowledgment}
This work was supported by the Laboratory Directed Research and Development program at Sandia National Laboratories, a multimission laboratory managed and operated by National Technology and Engineering Solutions of Sandia, LLC, a wholly owned subsidiary of Honeywell International Inc., for the U.S. Department of Energy’s National Nuclear Security Administration under contract DE-NA0003525. This work was performed, in part, at the Center for Integrated Nanotechnologies, an Office of Science User Facility, operated for the US Department of Energy Office of Science. This paper describes objective technical results and analysis. Any subjective views or opinions that might be expressed in the paper do not necessarily represent the views of the U.S. Department of Energy or the United States Government.

\bmsection{Author Contributions}
M.J.S., P.T.R., N.T.O. and M.E. conceived of the device concepts and experimental implementations. M.J.S. and J.H.D. performed the experiments and data analysis of the main devices. M.M. fabricated the devices. S.A.D. and M.E. conceived of the oscillator concepts. K.H. and M.E. derived the theory and operational analysis of the oscillator concepts. M.J.S., J.H.D., K.H., N.T.O. and M.E. wrote the paper.

% ========================================================================== %
%                                 Disclosures                                %
% ========================================================================== %

\bmsection{Disclosures}
The authors declare no conflicts of interest.

% ========================================================================== %
%                         Data Availability Statement                        %
% ========================================================================== %

\bmsection{Data availability} Data underlying the results presented in this paper are not publicly available at this time but may be obtained from the authors upon reasonable request.

% ========================================================================== %
%                            Supplemental Document                           %
% ========================================================================== %

% \bmsection{Supplemental document}
% A supplemental document must be called out in the back matter so that a link can be included. For example, “See Supplement 1 for supporting content.” Note that the Supplemental Document must also have a callout in the body of the paper.

\end{backmatter}

% ========================================================================== %
% -------------------------------------------------------------------------- %
%                                References                                  %
% -------------------------------------------------------------------------- %
% ========================================================================== %

\bibliography{maeorefs}

\begin{thebibliography}{10}
\newcommand{\enquote}[1]{``#1''}

\bibitem{capmany2007microwave}
J.~Capmany and D.~Novak, \enquote{Microwave photonics combines two worlds,} {\protect\JournalTitle{Nature photonics}} \textbf{1}, 319 (2007).

\bibitem{marpaung2019integrated}
D.~Marpaung, J.~Yao, and J.~Capmany, \enquote{Integrated microwave photonics,} {\protect\JournalTitle{Nature photonics}} \textbf{13}, 80--90 (2019).

\bibitem{aryanfar2017chip}
I.~Aryanfar, D.~Marpaung, A.~Choudhary, \emph{et~al.}, \enquote{Chip-based brillouin radio frequency photonic phase shifter and wideband time delay,} {\protect\JournalTitle{Optics letters}} \textbf{42}, 1313--1316 (2017).

\bibitem{mckay2020integrated}
L.~McKay, M.~Merklein, Y.~Liu, \emph{et~al.}, \enquote{Integrated microwave photonic true-time delay with interferometric delay enhancement based on brillouin scattering and microring resonators,} {\protect\JournalTitle{Optics Express}} \textbf{28}, 36020--36032 (2020).

\bibitem{gertler2022narrowband}
S.~Gertler, N.~T. Otterstrom, M.~Gehl, \emph{et~al.}, \enquote{Narrowband microwave-photonic notch filters using brillouin-based signal transduction in silicon,} {\protect\JournalTitle{Nature Communications}} \textbf{13}, 1947 (2022).

\bibitem{yao2002brillouin}
X.~S. Yao, \enquote{Brillouin selective sideband amplification of microwave photonic signals,} {\protect\JournalTitle{IEEE Photonics Technology Letters}} \textbf{10}, 138--140 (2002).

\bibitem{erdil2024wideband}
M.~Erdil, Y.~Deng, Z.~Tang, \emph{et~al.}, \enquote{Wideband, efficient alscn-si acousto-optic modulator in a commercially available silicon photonics process,} {\protect\JournalTitle{arXiv preprint arXiv:2402.01127}}  (2024).

\bibitem{tadesse2014sub}
S.~A. Tadesse and M.~Li, \enquote{Sub-optical wavelength acoustic wave modulation of integrated photonic resonators at microwave frequencies,} {\protect\JournalTitle{Nature communications}} \textbf{5}, 5402 (2014).

\bibitem{zhang2024integrated}
L.~Zhang, C.~Cui, P.-K. Chen, and L.~Fan, \enquote{Integrated-waveguide-based acousto-optic modulation with complete optical conversion,} {\protect\JournalTitle{Optica}} \textbf{11}, 184--189 (2024).

\bibitem{shao2020integrated}
L.~Shao, N.~Sinclair, J.~Leatham, \emph{et~al.}, \enquote{Integrated microwave acousto-optic frequency shifter on thin-film lithium niobate,} {\protect\JournalTitle{Optics Express}} \textbf{28}, 23728--23738 (2020).

\bibitem{wang2013optical}
X.~Wang, E.~H. Chan, and R.~A. Minasian, \enquote{Optical-to-rf phase shift conversion-based microwave photonic phase shifter using a fiber bragg grating,} {\protect\JournalTitle{Optics Letters}} \textbf{39}, 142--145 (2013).

\bibitem{tian2021magnetic}
H.~Tian, J.~Liu, A.~Siddharth, \emph{et~al.}, \enquote{Magnetic-free silicon nitride integrated optical isolator,} {\protect\JournalTitle{Nature Photonics}} \textbf{15}, 828--836 (2021).

\bibitem{sohn2021electrically}
D.~B. Sohn, O.~E. {\"O}rsel, and G.~Bahl, \enquote{Electrically driven optical isolation through phonon-mediated photonic autler--townes splitting,} {\protect\JournalTitle{Nature Photonics}} \textbf{15}, 822--827 (2021).

\bibitem{peterson2018synthetic}
C.~W. Peterson, S.~Kim, J.~T. Bernhard, and G.~Bahl, \enquote{Synthetic phonons enable nonreciprocal coupling to arbitrary resonator networks,} {\protect\JournalTitle{Science Advances}} \textbf{4}, eaat0232 (2018).

\bibitem{white1962amplification}
D.~L. White, \enquote{Amplification of ultrasonic waves in piezoelectric semiconductors,} {\protect\JournalTitle{Journal of Applied Physics}} \textbf{33}, 2547--2554 (1962).

\bibitem{carleton1965ultrasonic}
H.~Carleton, H.~Kroger, and E.~Prohofsky, \enquote{Ultrasonic effects in piezoelectric semiconductors,} {\protect\JournalTitle{Proceedings of the IEEE}} \textbf{53}, 1452--1464 (1965).

\bibitem{white1967surface}
R.~White, \enquote{Surface elastic-wave propagation and amplification,} {\protect\JournalTitle{IEEE Transactions on Electron Devices}} \textbf{14}, 181--189 (1967).

\bibitem{hackett2019amp}
L.~Hackett, A.~Siddiqui, D.~Dominguez, \emph{et~al.}, \enquote{High-gain leaky surface acoustic wave amplifier in epitaxial \uppercase{I}n\uppercase{G}a\uppercase{A}s on lithium niobate heterostructure,} {\protect\JournalTitle{Appl. Phys. Lett.}} \textbf{114}, 253503 (2019).

\bibitem{hackett2021towards}
L.~Hackett, M.~Miller, F.~Brimigion, \emph{et~al.}, \enquote{Towards single-chip radiofrequency signal processing via acoustoelectric electron--phonon interactions,} {\protect\JournalTitle{Nature communications}} \textbf{12}, 2769 (2021).

\bibitem{hackett2023non}
L.~Hackett, M.~Miller, S.~Weatherred, \emph{et~al.}, \enquote{Non-reciprocal acoustoelectric microwave amplifiers with net gain and low noise in continuous operation,} {\protect\JournalTitle{Nature Electronics}} \textbf{6}, 76--85 (2023).

\bibitem{storey2021acoustoelectric}
M.~J. Storey, L.~Hackett, S.~DiGregorio, \emph{et~al.}, \enquote{Acoustoelectric surface acoustic wave switch in an epitaxial ingaas on lithium niobate heterostructure,} in \emph{2021 21st International Conference on Solid-State Sensors, Actuators and Microsystems (Transducers),}  (IEEE, 2021), pp. 545--548.

\bibitem{wendt2025electrically}
A.~Wendt, M.~J. Storey, M.~Miller, \emph{et~al.}, \enquote{An electrically injected and solid state surface acoustic wave phonon laser,} {\protect\JournalTitle{arXiv preprint arXiv:2505.14385}}  (2025).

\bibitem{crowley1977acoustoelectric}
J.~Crowley, J.~Weller, and T.~Giallorenzi, \enquote{Acoustoelectric saw phase shifter,} {\protect\JournalTitle{Applied Physics Letters}} \textbf{31}, 558--560 (1977).

\bibitem{pedros2010voltage}
J.~Pedros, F.~Calle, R.~Cuerdo, \emph{et~al.}, \enquote{Voltage tunable surface acoustic wave phase shifter on algan/gan,} {\protect\JournalTitle{Applied Physics Letters}} \textbf{96} (2010).

\bibitem{yamagata2022surface}
M.~Yamagata, N.~Cao, D.~D. John, and H.~Hashemi, \enquote{Surface-acoustic-wave waveguides for radio frequency signal processing,} {\protect\JournalTitle{IEEE Transactions on Microwave Theory and Techniques}} \textbf{71}, 931--944 (2022).

\bibitem{hackett2024giant}
L.~Hackett, M.~Koppa, B.~Smith, \emph{et~al.}, \enquote{Giant electron-mediated phononic nonlinearity in semiconductor--piezoelectric heterostructures,} {\protect\JournalTitle{Nature Materials}} \textbf{23}, 1386--1393 (2024).

\bibitem{zhou2024electrically}
Y.~Zhou, F.~Ruesink, M.~Pavlovich, \emph{et~al.}, \enquote{Electrically interfaced brillouin-active waveguide for microwave photonic measurements,} {\protect\JournalTitle{Nature Communications}} \textbf{15}, 6796 (2024).

\bibitem{katzman2021surface}
M.~Katzman, D.~Munk, M.~Priel, \emph{et~al.}, \enquote{Surface acoustic microwave photonic filters in standard silicon-on-insulator,} {\protect\JournalTitle{Optica}} \textbf{8}, 697--707 (2021).

\bibitem{munk2019surface}
D.~Munk, M.~Katzman, M.~Hen, \emph{et~al.}, \enquote{Surface acoustic wave photonic devices in silicon on insulator,} {\protect\JournalTitle{Nature communications}} \textbf{10}, 4214 (2019).

\bibitem{gao2024compact}
L.~Gao, Y.~Liang, J.~Chen, \emph{et~al.}, \enquote{Compact low-half-wave-voltage thin film lithium niobate electro-optic phase modulator fabricated by photolithography-assisted chemo-mechanical etching (place),} {\protect\JournalTitle{Optics Letters}} \textbf{49}, 5783--5786 (2024).

\bibitem{ren2019integrated}
T.~Ren, M.~Zhang, C.~Wang, \emph{et~al.}, \enquote{An integrated low-voltage broadband lithium niobate phase modulator,} {\protect\JournalTitle{IEEE photonics technology letters}} \textbf{31}, 889--892 (2019).

\bibitem{Shin_Cox_Jarecki_Starbuck_Wang_Rakich_2015}
H.~Shin, J.~A. Cox, R.~Jarecki, \emph{et~al.}, \enquote{Control of coherent information via on-chip photonic–phononic emitter–receivers,} {\protect\JournalTitle{Nature Communications}} \textbf{6}, 6427 (2015).

\bibitem{896145}
I.~Kuznetsova, B.~Zaitsev, S.~Joshi, and I.~Borodina, \enquote{Investigation of acoustic waves in thin plates of lithium niobate and lithium tantalate,} {\protect\JournalTitle{IEEE Transactions on Ultrasonics, Ferroelectrics, and Frequency Control}} \textbf{48}, 322--328 (2001).

\bibitem{Xie_Sun_Peng_Zhang_Guo_Zhu_Hu_Chen_2014}
X.~Xie, T.~Sun, H.~Peng, \emph{et~al.}, \enquote{Low-noise and broadband optical frequency comb generation based on an optoelectronic oscillator,} {\protect\JournalTitle{Optics Letters}} \textbf{39}, 785 (2014).

\bibitem{Yao_Maleki_1996}
X.~S. Yao and L.~Maleki, \enquote{Optoelectronic microwave oscillator,} {\protect\JournalTitle{Journal of the Optical Society of America B}} \textbf{13}, 1725 (1996).

\bibitem{qi2020integrated}
Y.~Qi and Y.~Li, \enquote{Integrated lithium niobate photonics,} {\protect\JournalTitle{Nanophotonics}} \textbf{9}, 1287--1320 (2020).

\bibitem{kittlaus2021electrically}
E.~A. Kittlaus, W.~M. Jones, P.~T. Rakich, \emph{et~al.}, \enquote{Electrically driven acousto-optics and broadband non-reciprocity in silicon photonics,} {\protect\JournalTitle{Nature Photonics}} \textbf{15}, 43--52 (2021).

\bibitem{kenning2025broadband}
S.~E. Kenning, T.-H. Chang, A.~G. Attanasio, \emph{et~al.}, \enquote{Broadband acousto-optic modulators on silicon nitride,} {\protect\JournalTitle{Nature Communications}} \textbf{17}, 897 (2025).

\bibitem{Hassanien2021LNrel}
A.~E. Hassanien, S.~Link, Y.~Yang, \emph{et~al.}, \enquote{Efficient and wideband acousto-optic modulation on thin-film lithium niobate for microwave-to-photonic conversion,} {\protect\JournalTitle{Photon. Res.}} \textbf{9}, 1182--1190 (2021).

\bibitem{Kharel_Behunin_Renninger_Rakich_2016}
P.~Kharel, R.~O. Behunin, W.~H. Renninger, and P.~T. Rakich, \enquote{Noise and dynamics in forward brillouin interactions,} {\protect\JournalTitle{Physical Review A}} \textbf{93}, 063806 (2016).

\bibitem{Gertler_Kharel_Kittlaus_Otterstrom_Rakich_2020}
S.~Gertler, P.~Kharel, E.~A. Kittlaus, \emph{et~al.}, \enquote{Shaping nonlinear optical response using nonlocal forward brillouin interactions,} {\protect\JournalTitle{New Journal of Physics}} \textbf{22}, 043017 (2020).

\bibitem{250392}
M.~Kourogi, K.~Nakagawa, and M.~Ohtsu, \enquote{Wide-span optical frequency comb generator for accurate optical frequency difference measurement,} {\protect\JournalTitle{IEEE Journal of Quantum Electronics}} \textbf{29}, 2693--2701 (1993).

\end{thebibliography}

% ========================================================================== %
% -------------------------------------------------------------------------- %
%                            Attached Supplement                             %
% -------------------------------------------------------------------------- %
% ========================================================================== %

\appendix
% Title ==================================================================== %
\newpage
\title{Acoustoelectrically enhanced acousto-optic modulation in an integrated silicon nitride and thin film lithium niobate platform: supplemental document}

% ========================================================================== %
% -------------------------------------------------------------------------- %
%                     Acoustic Propagation Analysis                          %
% -------------------------------------------------------------------------- %
% ========================================================================== %
\section{Surface Acoustic Wave Simulations}

The acoustic wave properties are found using FEM simulations in COMSOL for the Rayleigh and SH surface acoustic waves. The acoustoelectric gain coefficient is proportional to the electromechanical coupling, $k^2$, when at a fixed-operating point. The electromechancical coupling, $k^2$ is extracted from simulation by comparing the difference in acoustic velocities when the top surface is electrically grounded, determining the interaction strength with the charge carriers in the InGaAs. Due to the anisotropic nature of lithium niobate, the coupling strongly depends on propagation angle, reaching zero for certain directions, and on the wavelength to LN thickness ratio. Hence, it is crucial to choose an ideal acoustic mode, angle of propagation, and acoustic wavelength. The fabricated devices target operation for the SH mode with a propagation angle along the crystal $+Y$-direction where the $k^2 \approx 10\%$. For different desired operating frequencies, the performance can be optimized by tuning the LN thickness and acoustic propagation angle. 

\begin{figure*}[ht]
    \centering
    \includegraphics[width=1\linewidth]{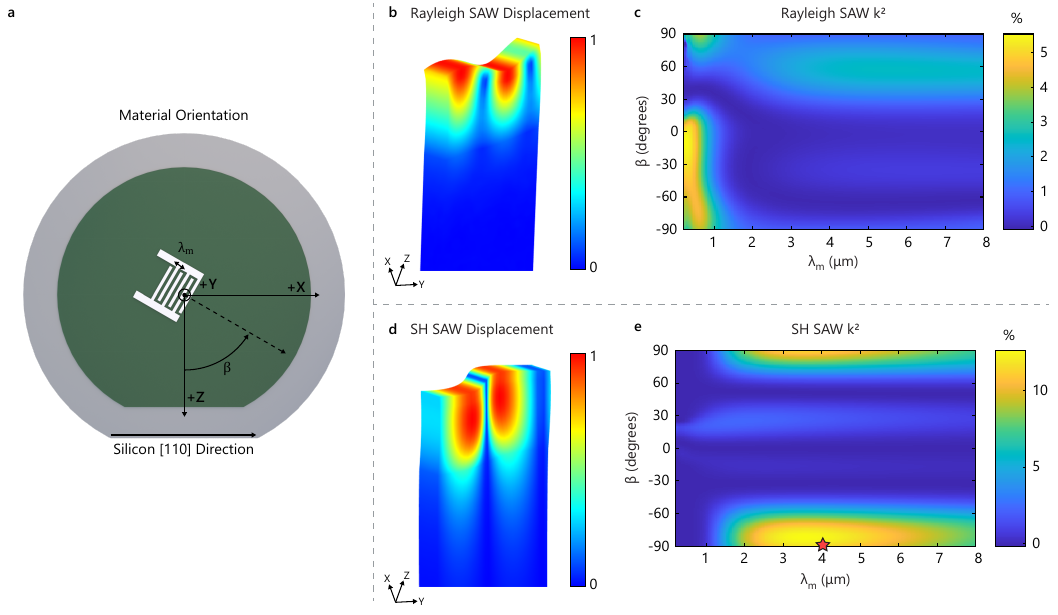}
    \caption{a) Orientation of the X-cut LN film with the propagation angle $\beta$ defined from +Z. b) and d) FEM simulations of displacement profiles for the respective Rayleigh and SH mode profiles. c) and d) Corresponding simulations of electromechanical coupling, $k^2$, as a function of propagation angle, $\beta$, and acoustic wavelength, $\lambda_m$. The electromechanical coupling is calculated by comparing acoustic wave velocity when the top surface of the LN has open and shorted electrical boundaries.}
    \label{fig:acoustic-sim}
\end{figure*}

% ========================================================================== %
% -------------------------------------------------------------------------- %
%                Acoustoelectric Performance Characterization                %
% -------------------------------------------------------------------------- %
% ========================================================================== %
\section{Acoustoelectric Performance Characterization}

\subsection{Acoustoelectric Experimental Setup}
The experimental setup consists of a VNA as an RF electronic source used to generate SAW with an interdigital transducer (IDT). The acoustic wave travels through the acoustoelectric gain medium with the drift current continuously provided by a DC voltage from an SMU. The SMU also measures the current through the InGaAs. A second IDT converts acoustic signal to electronic signal for direct readout of the S21 measurements on the VNA. All of the following measurements are taken in this configuration.

\begin{figure*}[ht]
    \centering
    \includegraphics[width=1\linewidth]{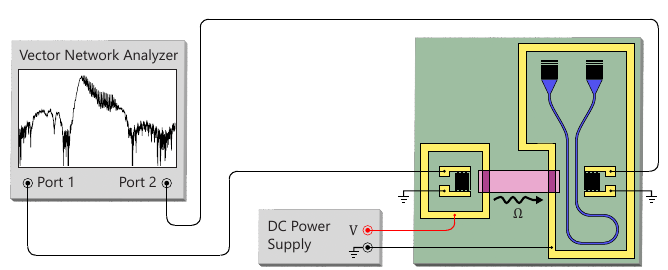}
    \caption{Experimental schematic used to take S21 measurements for the acoustic delay line with an acoustolectric amplifier. The VNA provides an RF source and measures S21 transmission between two IDTs. The DC voltage across the InGaAs provides a drift current for the AE interaction.}
    \label{fig:vna-setup}
\end{figure*}

\subsection{Baseline IL Measurement}
A baseline measurement of the S21 parameters is taken without any epitaxial material present. This accounts for the losses in the conversion from electronic to acoustic signals in the IDTs, as well as any intrinsic losses of the SAW from material losses or scattering off the SiN optical waveguide. This baseline measurement is roughly equivalent to a device of the same length with the epitaxial layers operated at the equal velocity point, where the carrier drift velocity equals the acoustic phase velocity. Here the AE effect provides no additional gain or attenuation, but there is some mass loading and scattering from the epi layer. Still, the baseline IL measurement allows determination of the increased performance of including the acoustoelectric amplifier, as well as the added losses of including the epi in passive operation. The baseline insertion loss of the SH mode is $-17$dB.

\begin{figure*}[ht]
    \centering
    \includegraphics[width=1\linewidth]{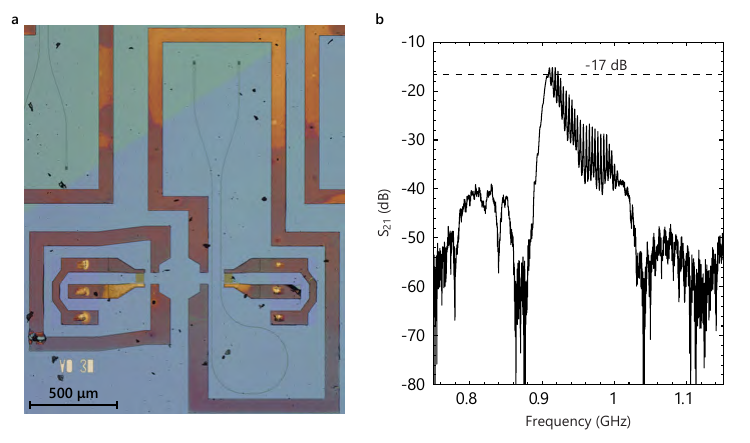}
    \caption{a) Device with no epitaxial layer which provides a baseline IL measurement of the S21 spectrum shown in b). The peak at -17dB indicates the frequency at which the IDTs are well phase-matched to excite the SH mode.}
    \label{fig:IL-baseline}
\end{figure*}

\subsection{Acoustoelectric Strength vs. Applied DC Voltage}
\begin{figure*}[ht]
    \centering
    \includegraphics[width=1\linewidth]{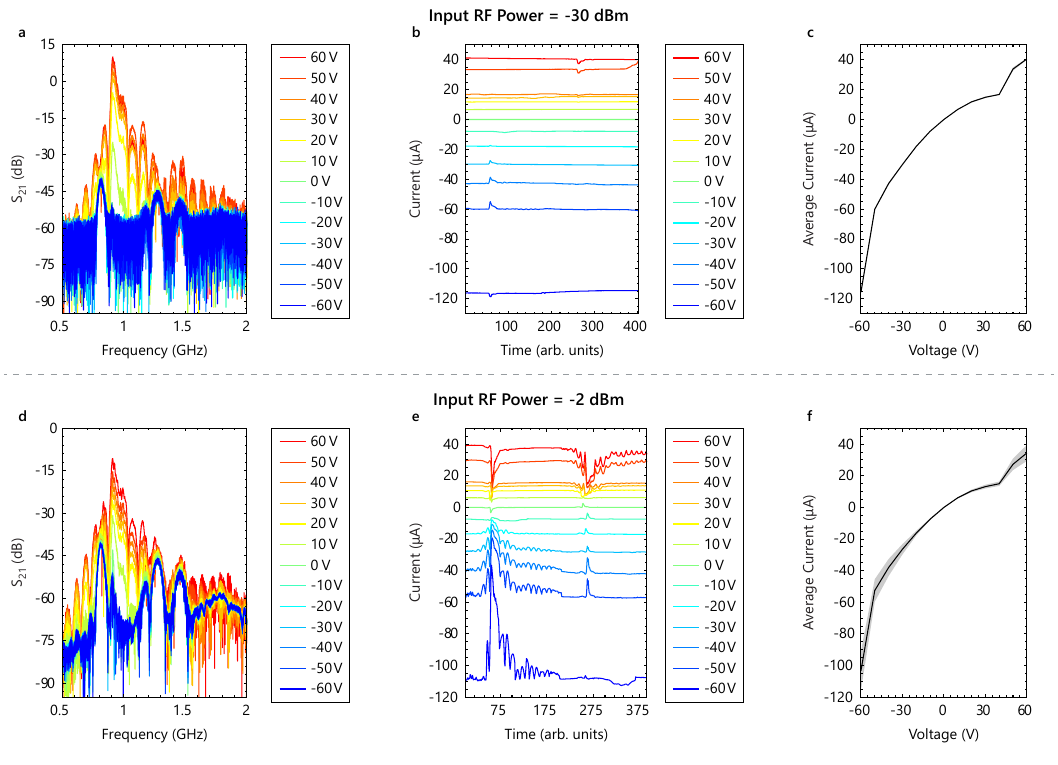}
    \caption{a) and d) Transmission spectrum for two input RF powers, -30dBm and -2 dBm, respectively, as the DC voltage is changed from -60V to +60V. As the drive frequency is swept in time, the current for a given voltage is measured b) and e). Between acoustic excitations the current is flat at a bias-dependent baseline, with the average current and standard deviation shown in c) and f), highlighting the non-Ohmic behavior of the contacts.}
    \label{fig:vna-sweeps}
\end{figure*}

The induced gain/loss from the acoustoelectric effect depends on the carrier drift velocity controlled with the applied DC voltage. The frequency dependence of the change in S21 with the applied DC voltage shown in Figure \ref{fig:vna-sweeps} highlights the strong electromechanical coupling of the SH mode. The Rayleigh mode at ~800MHz has a weak dependence of S21 with the DC voltage due to its comparatively weak $k^2$ at the fabricated propagation angle. The SH mode, centered near 900MHz, demonstrates the net gain of the AE amplifier where the AE gain allows for acoustic power greater than the IL measurement (i.e. net gain). 

\subsection{Acoustoelectric Strength vs. Input RF Power}
\begin{figure*}[ht]
    \centering
    \includegraphics[width=1\linewidth]{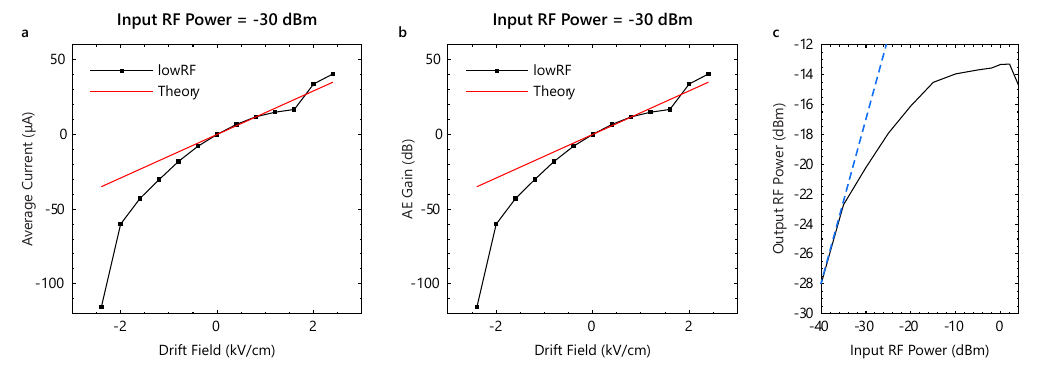}
    \caption{(a) Average current as a function of applied drift field fit with theory. (b) Measured acoustoelectric gain as a function of applied drift field fit with theory. Both fits show nonlinear behavior in the experimental data, which is a result of non-Ohmic contacts to the InGaAs thin film. (c) Output vs. Input RF power with -60V applied showing saturation effects in the acoustoelectric amplification. Linearity in the acoustoelectric gain is observed up to an input RF power of -35dBm (dashed blue line). Larger input RF powers experience gain roll off.}
    \label{fig:vna-linedata}
\end{figure*}

For weak signals, the AE amplifier provides a fixed gain, however, for strong input RF powers, the acoustic amplitude is strong enough to saturate the carriers in the InGaAs. The S21 and DC current across the epitaxial layer are measured as the RF drive frequency is swept in time, with the S21 data shown in Figure \ref{fig:vna-sweeps} a and b. Nonlinear gain is clear in the fact that for -30dBm the amplifier achieves terminal gain ($S_{21} > 0$dB, but for the input power of -2dBm the amplifier achieves slight net gain. In Figure \ref{fig:vna-linedata}, the input RF power vs. ouput RF power for a fixed DC bias voltage shows the highly nonlinear gain. 

The gain saturation can be interpreted from the current across the InGaAs epitaxial layer. The current as function of measurement time is shown in Figures \ref{fig:vna-sweeps} b and e. The changes in current from the baseline value is from excited acoustic wave interactions with the charge carriers. The positive upticks in current occur at voltages where the carrier drift velocity lags the acoustic phase front and the acoustic wave accelerates carriers. Likewise, a negative change in current indicates acoustoelectric gain as energy is transferred from the drifting carriers to the acoustic wave, meaning the carrier drift velocity surpasses the acoustic waves' phase front. The deeper the dip in current, the larger fraction of drifting charge carriers are losing energy to amplify the acoustic wave. As the dip in current approach zero, the number of drifting carriers that can contribute to gain are depleted and the gain is clamped.

% ========================================================================== %
% -------------------------------------------------------------------------- %
%                     Acousto-optic Coupling and Design                      %
% -------------------------------------------------------------------------- %
% ========================================================================== %
\section{Acousto-optic Coupling and Design}

\subsection{Calculate AOM Efficiency, Vpi, VpiL}
To calculate the AOM efficiency by way of the $V_{\pi}$ we assume the optical input has the form $E_{in}=E_0 e^{i \omega t}$. After phase modulation, the resulting signal is $E_{out}=E_0 e^{i \omega t+i \frac{\pi V}{V_\pi} \sin(\Omega t)}$, where $V$ is the voltage of the applied RF signal and $\Omega$ is the modulation frequency. According to the Jacobi-Anger expansion, the modulated signal is
\begin{align}\label{Jacobi}
E_{out}=E_0 e^{i \omega t}\bigl[J_0(0)+2\sum_{n=1}^\infty i^n J_n(\frac{\pi V_{RF}}{V_\pi})\cos(n \Omega t)\bigr]
\end{align}
where $n=0$ is the carrier term and $n>0$ refer to the ordered side bands of the signal. To calculate the $V_\pi$ for our modulator, we define the AOM efficiency, $\eta$, as the ratio of modulated sideband power ($P_1$) to the carrier power ($P_0$) 
\begin{align}\label{efficiency}
\eta\equiv\frac{P_1}{P_0}=\frac{|J_1(\frac{\pi V_{RF}}{V_\pi})|^2}{|J_0(0)|^2}\approx\Bigl(\frac{\pi V_{RF}}{V_\pi}\Bigr)^2,
\end{align}
where the last step makes the small modulation approximation.

% ========================================================================== %
% -------------------------------------------------------------------------- %
%                            Supplement AE-OAO                               %
% -------------------------------------------------------------------------- %
% ========================================================================== %

\section{Small-signal threshold condition of AE-OAO}
In the following analysis, we describe the pertinent physics to determining when the acoustoelectric opto-acoustic oscillator (AE-OAO) reaches threshold. A noise-seeded signal will regenerate itself when the round-trip gain is greater than unity. The round-trip signal path is given as,
\begin{equation}
    x \xrightarrow{\text{Receiver resonator}} \{A_n^{(b)} \} \xrightarrow{\text{coupler}}
    A_{bus} \xrightarrow{\text{ring}}
    A_{bus}'\xrightarrow{\text{coupler}}\{A_n^{(a)} \} \xrightarrow{\text{Emitter resonator}} 
     S(\Omega)
    \xrightarrow{\chi_m g^* G^{1/2}_{AE}} x_{out}
\end{equation}. 
\begin{table}[]
    \centering
    \begin{tabular}{|c|c|}
      \hline
      $x$   &  Modulation depth and phase\\
      $A$   &  Optical field amplitude\\
      $B$   & Acoustic field amplitude\\
      $\omega$ & Optical carrier frequency \\
      $\Omega$ & Acoustic frequency/Comb repetition rate \\
      $\Gamma$ & Acoustic energy loss rate\\
      $g$   & Brillouin optomechanical coupling \\
      $\rho$ & Round-trip transmission of optical resonators\\
      $r,t$ & Coupler reflection/transmission coefficients \\
      $G_{AE}$ & Net acoustoelectric gain \\
      \hline
\end{tabular}
    \caption{Variable definitions}
    \label{tab:placeholder}
\end{table}

The optical field in the emitter (a) waveguide can be expanded as a superposition of waves with arbitrary phase and amplitude spaced by frequency $\Omega$ at $z=0$,
\begin{align}
    A^{(a)}|_{z=0} = \sum_{n=-\infty}^\infty A_n e^{-i(\omega  + \Omega n)t}
\end{align}
Further, an acoustic mode is supported in the waveguide, inducing scattering between optical modes when the contributing fields satisfy phase-matching constraints. The device will focus on interactions between forward propagating optical modes in the same band (i.e. forward intramodal SBS). In this limit, the acoustic phase-velocity parallel to the optical propagation direction must be equal to the optical group velocity. These conditions are satisfied a phonon mode that has near-zero group velocity along the optical propagation direction. Further, a cascade of scattering can occur between optical modes in the zero group velocity limit, allowing generation of frequency combs. The interaction Hamiltonian betwen the optical modes and acoustic mode is,
\begin{align}
    H = \hbar \sum_n (g^* A^\dagger_{n+1} A_n B + g A^\dagger_{n} A_{n+1} B^\dagger) 
\end{align}
summing over the phase-matched optical modes.
In general, the spatial evolution of each mode in the identical emitter and receiver can be described by the following equations of motion \cite{Kharel_Behunin_Renninger_Rakich_2016},
\begin{align}
    \frac{\partial A_n}{\partial t} + v_n  \frac{\partial A_n}{\partial z} =-\kappa A_n - i(g A_{n+1} B^\dagger + g
    ^* B A_{n-1})
    \label{eq: opt_eom}
\end{align}
\begin{align}
    \frac{\partial B}{\partial t} + v_0  \frac{\partial B}{\partial z} = i (\Omega - \Omega_0) B - \frac{\Gamma_0}{2} B - ig \sum_{n=-\infty}^\infty A_{n}^\dagger A_{n+1}
    \label{eq: aco_eom}
\end{align}
We are primarily interested in the steady-state solution of the system, hence, the time-derivatives are set to zero. Further, under the condition $v_{ac}/ \Gamma_0 \ll L_{pm}$, the acoustic wave then acts like a Raman scatterer, only interacting locally each point along z. Therefore, the spatial derivative of the acoustic equation of motion is ignored because the near-zero group velocity and high acoustic decay rate prevents a build up of acoustic field along the length of the device. The simplified equations of motion are,
\begin{align}
   v_n  \frac{\partial A_n}{\partial z} =-\kappa A_n - i(g A_{n+1} B^\dagger + g
    ^* B A_{n-1})
    \label{eq: opt_eom_simp}
\end{align}
\begin{align}
    B(z) = i\chi_m g \sum_{n=-\infty}^\infty A_{n}^\dagger(z) A_{n+1}(z)
    \label{eq: aco_eom_simp}
\end{align}
where $\chi_m = \sfrac{1}{(\Gamma_0/2 - i(\Omega - \Omega_0))}$ is the mechanical susceptibility. It is important to note that in Equation \ref{eq: aco_eom_simp}, the optical drive term shows the necessity of amplitude modulation. If there is amplitude modulation, this term will be nonzero and the acoustic field will be seeded. Interestingly, the optical fields will slowly be red-shifted along the length of the device, due to the back-action of the acoustic fields on the optical field \cite{Gertler_Kharel_Kittlaus_Otterstrom_Rakich_2020}. However, the acoustic waves only generate phase modulation of the optical field, never changing the amplitude modulation term. Here the acoustic mode is assumed to be a leaky mode and it strongly couples to a surface acoustic wave that radiates outwards. The acoustic dissipation rate is expanded as,
\begin{align}
    \Gamma_0 = \Gamma_{int} + \Gamma_{bulk} + \Gamma_{surf,+} + \Gamma_{surf,-}
\end{align}
where $\Gamma_{int}$ is the intrinsic losses due to the material, $\Gamma_{bulk}$ is the loss into the bulk modes of the layer, $\Gamma_{surf,+}$ is the loss into the surface acoustic waves that propagates to the left and $\Gamma_{surf,-}$ is the loss into surface acoustic wave that propagates towards the receiver waveguide. Considering the acoustic dissipation and phase accumulation of the acoustic wave across the acoustoelectric gain region, the expression for the acoustic wave at the receiver waveguide is,
\begin{align}
    B_{b}(z) = \chi_m \sqrt{\Gamma_{surf,-}\Gamma_{surf,+}} e^{\alpha L_g/2} e^{i \phi} B_a(z)
    \label{eq: aco_eom_rec_2}
\end{align}
The acoustic wave in the receiver waveguide is simply a constant transfer function of the acoustic field in the emitter waveguide. This acoustic wave will then act on the optical wave in the reveiver waveguide, where thee optical wave dynamics are given by Equation $\ref{eq: opt_eom_simp}$. In the limit of zero GVD, then $v_n = v$ for all relevant $n$. Also, the optomechanical coupling $g$ is assumed constant across all inter-frequency terms. Defining, $\psi = \arg(g^*B^{(b)})$, the full optical field solution is,
\begin{align}
    A(z) = A(0) \exp[-i \zeta z \cos(\Omega t - \psi)] e^{-\kappa z/v}
\end{align}
with $\zeta = 2|gB|/v$ as the modulation depth per unit length. In order to enhance the optomechanical interactions, the receiver waveguide is designed to be an optical resonator with a free-spectral range of the desired oscillation frequency. This produces an optical frequency comb identical to a Kourogi comb, where a phase-modulator is enhanced by a resonant structure \cite{250392}. The resonantly enhanced field is given as,
\begin{align}
    A^{(b)}(t) = ir_b a_{in} \sum_{k=0}^{\infty} (\rho_b t_b)^k \exp[-i k\zeta L_{pm} \cos(\Omega t - \psi)]
\end{align}
This optical field produces pulses at a $2\Omega$ frequency for small modulation depths, $|x| = \zeta L_{pm} < \pi$. However, to generate acoustic modes in the emitter waveguide, there must be an intensity signal at frequency $\Omega$. A dispersive element is required to shift the relative phases of the sidebands to generate an amplitude modulated signal. In the small-signal limit only the first-order sidebands are relevant. We assume that total phase to amplitude modulation is achieved, which is possible for example with an overcoupled ring resonator. If the coupling rate $\kappa_{ext} \ll \kappa_{int} \hspace{1mm} \& \hspace{1mm}  \kappa_{ext} + \kappa_{int} < \Omega$, then if the optical carrier frequency is detuned by half the loaded linewidth, the relative phase of the carrier will be shifted by $\pi/2$ to the sidebands, converting the phase-modulated signal to pure amplitude modulation. Assuming unity conversion, the small-signal round-trip gain is given as,
\begin{align}
    G_{s} = C \left( t_b - \frac{r_b^2 \rho_b}{ 1 - t_b\rho_b} \right) \left(\frac{ t_b \rho_b}{2(1 - t_b\rho_b)^2} \right)
\end{align}
with 
\begin{align}
    C(\omega) = \frac{2i\chi_m^2(\omega) |g(\omega)|^2 G^{1/2}_{ae} L_{pm}\sqrt{\Gamma_{surf,-}\Gamma_{surf,+}}|s_{in}|^2 \rho_b e^{-i\Omega T_a}}{v}\frac{ r_b^2 r_a^2}{(1-t_a \rho_{a})^2}
\end{align}

\end{document}